\documentclass[pdftex,twocolumn,epjc3]{svjour3}          

\RequirePackage[T1]{fontenc}

\smartqed  

\RequirePackage{graphicx}
\RequirePackage{mathptmx}      
\RequirePackage{flushend}
\RequirePackage[numbers,sort&compress]{natbib}
\RequirePackage[colorlinks,citecolor=blue,urlcolor=blue,linkcolor=blue]{hyperref}

\usepackage{amsmath,amssymb,bm}
\usepackage{float}
\usepackage[normalem]{ulem}

\usepackage{color}

\journalname{Eur. Phys. J. B}

\begin{document}

\title{Three-terminal normal-superconductor junction as thermal transistor}

\author{Gaomin Tang\thanksref{e1,addr1}
        \and
        Jiebin Peng\thanksref{addr2} 
        \and
        Jian-Sheng Wang\thanksref{addr1}}

\thankstext{e1}{e-mail: gmtang1212@gmail.com}

\institute{Department of Physics, National University of Singapore, Singapore 117551, Republic of Singapore\label{addr1}
          \and
Center for Phononics and Thermal Energy Science, China-EU Joint Center for Nanophononics, Shanghai Key Laboratory of Special Artificial Microstructure Materials and Technology, School of Physics Science and Engineering, Tongji University, 200092 Shanghai, China\label{addr2} }

\date{Received: date / Accepted: date}

\maketitle

\begin{abstract}
We propose a thermal transistor based on a three-terminal normal-superconductor (NS) junction with superconductor terminal acting as the base.
The emergence of heat amplification is due to the negative differential thermal conductance (NDTC) effect for the NS diode in which the normal side maintains a higher temperature. 
The temperature dependent superconducting energy gap is responsible for the NDTC. 
By controlling quantum dot levels and their coupling strengths to the terminals, a huge heat amplification factor can be achieved. 
The setup offers an alternative tuning scheme of heat amplification factor and may find use in cryogenic applications. 
\end{abstract}

\section{Introduction}
Thermal management at nanoscale is receiving significant research attention \cite{RMP_Li}, and there have been a variety of topics on this subject: spin caloritronics \cite{spin-calori}, near-field radiative heat transfer \cite{near-field,near-field1}, energy harvesting \cite{harvester}, cooling \cite{cooling,cooling-atom, refrigeration,CPS-cooling}, thermometry \cite{thermometry1,thermometry2,thermometry3}, heat-assisted magnetic recording \cite{heat-record}, and so on. As the diodes and transistors are the basic elements of electronic circuits, various thermal diodes \cite{RMP_Li, thermal_rectifier,QD_thermal_rectifier, Jie1, NFRHT_diode, NS_diode0, NS_diode1, Si_diode, gm1, gm2} and thermal transistors \cite{TT_Li1, TT_Li2, TT_11, TT_phonon, TT_NF, TT_FF1, TT_FF2, QTT, TT_JJ, TT_Rafa, TT_Coulomb, TT_Ren1, TT_Ren2} have been proposed in order to realize the solid-state thermal circuits. In a thermal diode, heat current changes the magnitude with the reversal of the temperature bias. Similar to an electronic transistor \cite{Bardeen, Shockley}, of which the collector electric current is amplified relatively to the base current [see Figure~\ref{fig1} (a)], the function of a thermal transistor is to modulate the heat current in the collector with a small change of the heat current in the base. Li {\it et al.} proposed the first model of all-thermal transistor using nonlinear lattices and uncovered that negative differential thermal conductance (NDTC) is crucial to realize heat amplification \cite{TT_Li1, TT_Li2}. The phenomenon of NDTC refers to an enhanced  heat current with decreasing temperature bias in a two-terminal setup \cite{NDTC1, NDTC2, NDTC3}.


	Compared to controlling electric currents, managing heat currents by means of changing temperature is much more complicated. Very recently, thermal transistor effect has been proposed in various nanostructures \cite{QTT, TT_JJ, TT_Rafa, TT_Coulomb, TT_Ren1, TT_Ren2}. In a setup with three interacting two-level subsystems based on spin-boson model, strong couplings between the two-level subsystems are needed to achieve huge heat amplification \cite{QTT, TT_Ren1}. Fornieri {\it et al.} reported NDTC and heat amplification in a temperature-biased Josephson tunnel junction between two different superconductors \cite{TT_JJ}. Thermal fluctuations from the base terminal are employed to induce huge amplification factors \cite{TT_Rafa}. Heat amplification in a Coulomb-coupled triple-quantum-dots setup can be obtained by tuning the Coulomb interaction between the collector and the emitter \cite{TT_Coulomb}. The realization of thermal transistor was also proposed in cavity-coupled double-quantum-dots system, in which the strong light-matter interactions are needed to achieve thermal amplification \cite{TT_Ren2}.

	Recently, superconductors, especially Josephson junctions have been applied to construct various thermal devices, such as thermal diode \cite{NS_diode0, NS_diode1}, heat interferometer \cite{heat_interfere}, heat modulator \cite{heat_modulate}, thermoelectric device \cite{SC_spincalori}, thermal router \cite{router}, thermal memory \cite{thermal_memory}, and the above mentioned thermal transistor in Josephson junction \cite{TT_JJ}. In this work, we investigate heat amplification using a three-terminal normal-superconductor (NS) junction [see Figure~\ref{fig1}(b)], by utilizing the possible NDTC effect in a two-terminal NS junction \cite{TIS_diode, NDTC-SC}. To achieve heat amplification, there is no need to have interactions in the hybrid junction explicitly. The temperature dependent superconducting energy gap due to electron-phonon interactions inside the superconductor \cite{BCS} is responsible for the occurrence of NDTC. The quantum dot levels and their couplings to the terminals provide alternative approaches to tune the amplification factor. The scheme of heat amplification using three terminal normal-superconductor junction can find its application in cryogenic thermal management. 

	The article is structured as follows. In Sec.~\ref{sec:II}, the model Hamiltonian and nonequilibrium Green's function formalism are given for the hybrid junction, and the mechanism for heat amplification is uncovered. Numerical results are shown in Sec.~\ref{sec:III} demonstrating huge amplification factors can be easily obtained. We finally summarize our work in Sec.~\ref{sec:IV}.

\section{Model and mechanism}  \label{sec:II}
\begin{figure}
\includegraphics[width=3.3in]{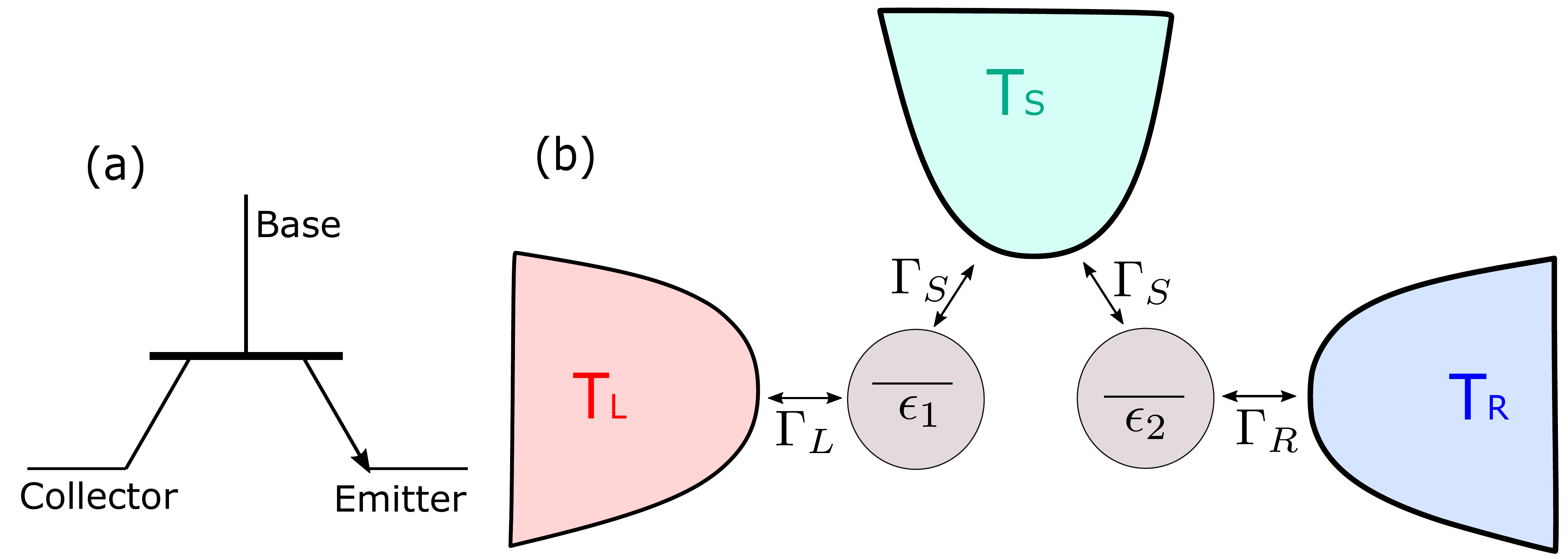} \\
\caption{(Color online) (a). A schematic symbol for the bipolar junction transistor. (b). A three-terminal junction is used to realize the thermal transistor. Two normal leads are maintained at temperatures $T_L$ and $T_R$, respectively. The superconducting lead has the temperature $T_S$ and acts as the base. Two quantum dots offer the tunability of the heat amplification. }
\label{fig1}
\end{figure}

The model we consider is sketched in Figure~\ref{fig1}(b) where two quantum dots (QDs) are placed between each normal-metal terminal and the superconductor. 
The total Hamiltonian of the system can be written as
\begin{equation}
H = H_{QDs} + H_L + H_R + H_S + H_T ,
\end{equation}
where the Hamiltonian of two QDs is 
\begin{equation}
H_{QDs} =\sum_{m,\sigma}\epsilon_{m} d_{m\sigma}^\dag d_{m\sigma} +U \hat{n}_{m\uparrow} \hat{n}_{m\downarrow},
\end{equation}
with QD levels $\epsilon_{m}$ and $m=1,2$. The second term of $H_{QDs}$ describes the Coulomb interaction in QDs with the number operator $\hat{n}_{m\sigma} = d_{m\sigma}^\dag d_{m\sigma}$, and we assume that both QDs have the same Coulomb interaction strength. Each QD has only one energy level with spin degeneracy, which is enough to capture the essential physics and facilitates experimental tuning as well. The two normal metal leads $\alpha=L,R$ are described by
\begin{equation}
H_{\alpha} = \sum_{k\sigma} \epsilon_{\alpha k} c_{\alpha k\sigma}^\dag c_{\alpha k\sigma} .
\end{equation}
The Hamiltonian of the s-wave superconducting lead is 
\begin{equation}
H_S =\sum_{k\sigma} \xi_k a_{k\sigma}^\dag a_{k\sigma} -\sum_k \left(\Delta a_{k\uparrow}^\dag a_{-k\downarrow}^\dag + {\rm H.c.} \right) ,
\end{equation} 
where we have ignored the superconducting phase and take $\Delta$ to be real and independent of momentum $k$. 
The superconducting energy gap $\Delta$ depends on the temperature of the superconductor and bears the gap interpolation formula \cite{gap1}
\begin{equation} \label{gap}
\Delta(T_S) = \Delta_0 \tanh\left[ \frac{\pi k_B T_c}{\Delta_0} \sqrt{a(T_c/T_S-1)} \right] , 
\end{equation}
with the zero temperature gap $\Delta_0$, critical temperature $T_c$ and Boltzmann constant $k_B$. The free parameter $a$ depends on the particular pairing state. 
The coupling Hamiltonian to the three leads is described by
\begin{align}
H_T = \sum_{k\sigma} \Big( & t_{Lk} c_{Lk\sigma}^\dag d_{1\sigma} 
+ t_{Sk} a_{k\sigma}^\dag d_{1\sigma}  \notag \\
+ & t_{Rk} c_{Rk\sigma}^\dag d_{2\sigma} 
+ t_{Sk} a_{k\sigma}^\dag d_{2\sigma} + {\rm H.c.} \Big) . 
\end{align}
Three terminals are at different temperatures $T_L$, $T_R$ and $T_S$, and no voltage bias is applied at each terminal. 

	The retarded self-energies due to the left and right normal lead are diagonal in Nambu space and assumed to be energy independent with $\bm{\Sigma}_{\alpha}^r =-i\bm{\Gamma}_\alpha /2$ where $\bm{\Gamma}_\alpha = {\rm diag}[\Gamma_\alpha, \Gamma_\alpha]$. The quantitative influence of band-widths of the leads on the thermal transistor behavior can be neglected in this work.
	 The retarded self-energies due to superconducting lead are assumed to be the same for both QDs and energy dependent with the form \cite{NS1, NS2}
\begin{equation}
\bm{\Sigma}^r_S (E) = - \frac{\Gamma_S}{ 2 \sqrt{|\Delta|^2-(E +i\eta_s)^2}}
\begin{pmatrix}
E +i\eta_s & -\Delta  \\  -\Delta  & E +i\eta_s
\end{pmatrix} ,
\end{equation}
where Dynes broadening parameter \cite{Dynes} $\eta_s$ is included. The effective linewidth function of superconducting lead is introduced as $\widetilde{\bm{\Gamma}}_S = i(\bm{\Sigma}^r_S - \bm{\Sigma}^a_S)$. 
Using the equation of motion approach and the mean-field decoupling scheme \cite{Pablo}, the retarded and advanced Green's functions ${\bf G}_m^{r/a}(E)$ for the site $m$ QD can be evaluated self-consistently (See Appendix for details).

\begin{figure*} 
\includegraphics[width=7.0in]{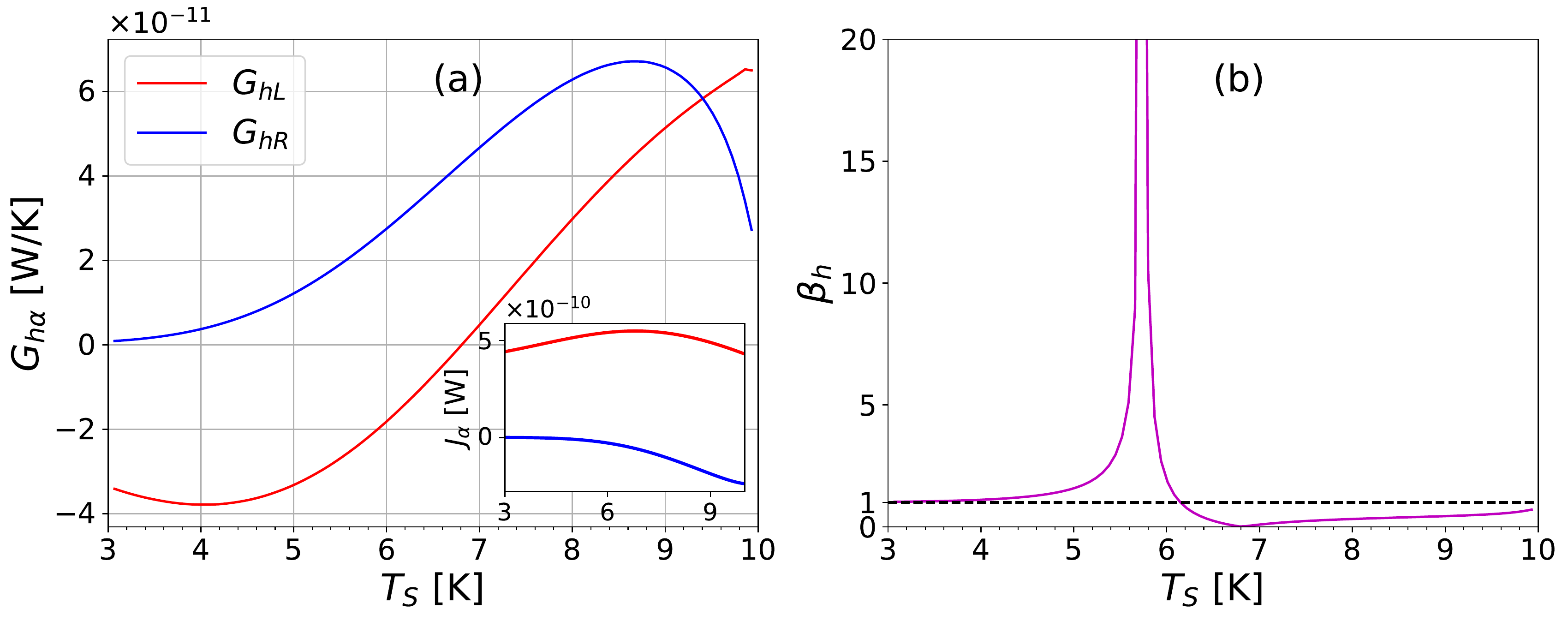} \\
\caption{(Color online) (a) Thermal conductance in the left and right normal lead, and (b) heat amplification factor, by varying superconductor's temperature within $T_S \in (T_R, T_c)$. QD levels are $\epsilon_1 =\epsilon_2 = 3.5\,$meV. Heat currents $J_\alpha$ are shown as the inset of panel (a).  }
\label{fig2}
\end{figure*}

	In the absence of voltage bias, heat current equals the corresponding energy current. Heat current in normal metal lead $\alpha$ is only induced by quasi-particle process (conventional electron tunneling) and expressed as \cite{SC_spincalori, NS2}:
\begin{equation} \label{heatcurrent}
J_{\alpha} = \frac{2}{\hbar} \int_{-\infty}^{\infty} \frac{dE}{2\pi} \ E T_{\alpha}^{qp}(E) \left[ f_{\alpha}(E)-f_S(E) \right],
\end{equation}
with the Fermi-Dirac distribution functions $f_{\alpha/S}(E)=$ \\ $ \{\exp[E/(k_B T_{\alpha/S})]+1 \}^{-1}$. 
The quasi-particle transmission coefficient for lead $\alpha$ reads as
\begin{align}
& T_{\alpha}^{qp}(E) = [\bm{\Gamma}_{\alpha} {\bf G}_m^r(E) \widetilde{\bm{\Gamma}}_S(E) {\bf G}_m^a(E) ]_{ee} ,
\end{align}
with $m=1(2)$ for $\alpha=L(R)$. The subscript `$ee$' in the above equation denotes that only the first entry of the matrix is taken. The total heat current flowing in superconducting lead is $J_{S} = -(J_L + J_R )$, thus fulfilling the energy current conservation law. We can see that the transmission coefficients for the left and right lead can be tuned separately by controlling the QD levels. 

	The heat amplification factor in our setup is quantitatively described by the change of heat current in the normal lead upon the change of that in the superconductor lead, which serves as the base. We can define the heat amplification with respect to the left lead in the form
\begin{equation} \label{define}
\beta_h^L = \left| \frac{\partial J_L}{\partial J_S} \right| = \left| \frac{\partial J_L}{\partial (J_L + J_R)} \right|
=\left| \frac{G_{hL}}{G_{hL} + G_{hR}} \right|. 
\end{equation}
$G_{h\alpha} = - \partial J_\alpha /\partial T_S |_{T_\alpha ={\rm const}}$ is the differential thermal conductance at lead $\alpha$ with respect to superconductor's temperature. Clearly, the necessary condition to realize heat amplification ($\beta_h^L >1$) is $G_{hL} G_{hR} <0$, that is, only one of $G_{hL}$ and $G_{hR}$ being negative. Therefore, NDTC in only one of the normal lead is necessary for the emergence of heat amplification. Beside this, $|G_{hR}|<2|G_{hL}|$ should be satisfied as well. Heat amplification factor can approach infinity with vanishing $G_{hL}+G_{hR}$, and this condition can be easily satisfied by tuning quantum dot levels as will be shown in the numerical section. For the occurring of heat amplification in the right lead, the conditions $G_{hL} G_{hR} <0$ and $|G_{hL}|<2|G_{hR}|$ should be satisfied. With $G_{hL} G_{hR} <0$, the condition for heat amplifications in both the left and right leads is $|G_{hR}|/2<|G_{hL}|<2|G_{hR}|$. The relation $\beta_h^L - \beta_h^R = \pm 1$ takes plus (minus) sign when NDTC happens in the left (right) lead. The above argument holds for a general thermal transistor.

	We now discuss the occurrence of NDTC in the case when the normal lead temperature is higher than the superconductor ($T_\alpha > T_S$). In this case, the heat current obtained from equation~\eqref{heatcurrent} is positive. With increasing superconductor temperature $T_S$, superconducting gap $\Delta(T_S)$ decreases, and this enhances the quasi-particle tunneling process, i.e., an increased $T_{\alpha}^{qp}$. However, the difference between Fermi-Dirac distributions $f_\alpha - f_S$ diminishes. The competition between the superconducting gap and Fermi-Dirac distribution difference offers the possibility to increase heat current by decreasing temperature difference, that is the NDTC effect. When the superconducting gap changing dominates over Fermi-Dirac distribution difference with varying superconductor temperature, NDTC effect emerges, otherwise, NDTC effect is absent \citep{NDTC-SC}.  The NDTC effect for normal-superconductor junctions have been reported in topological insulator superconductor junction \cite{TIS_diode} and graphene-based superconducting junction \cite{NDTC-SC}. 	
	In the case of $T_S > T_\alpha$, the two effects cooperate and NDTC is absent. Based on these facts, we can set the temperature profile as $T_L > T_S > T_R$. Then by tuning parameters (temperatures, dot levels, and dot-lead couplings), we can have NDTC effect between the left normal lead and the superconductor, and there is no such effect for the right lead. Then heat amplification is achieved. 
	One can also use a superconductor-normal metal-normal metal junction, wherein the central normal metal serves as the base, to realize the thermal transistor. The superconductor has the lowest temperature, and the base temperature is lower than the other normal terminal. Due to the presence of NDTC effect between the superconductor and base terminal and the absence of NDTC between two normal metals, heat amplification is achieved. 
	Using only one quantum dot can in principle achieve thermal transistor, however the tunability is lower compared to the case in this work.

\begin{figure*}
\includegraphics[width=7.0in]{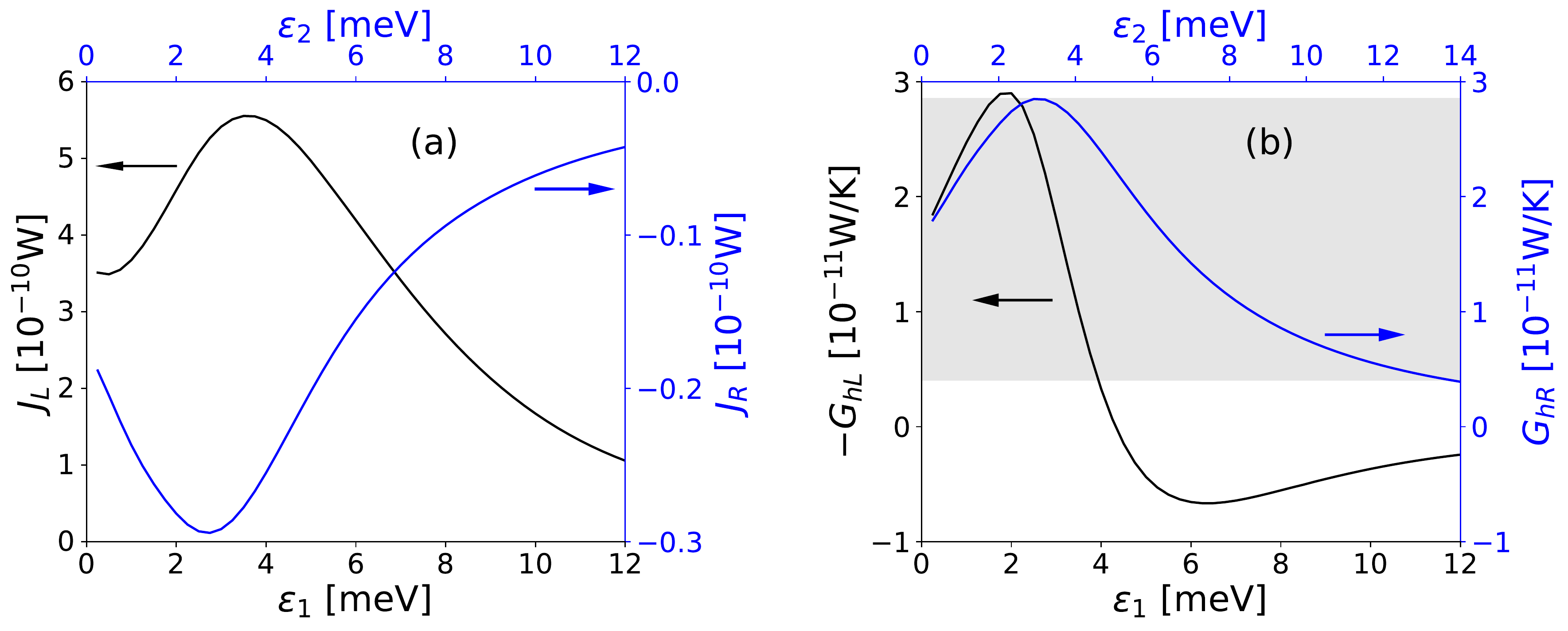} \\
\caption{(Color online) (a) Heat currents $J_\alpha$ and (b) thermal conductance $G_{h\alpha}$ with respect to the corresponding QD levels. The superconductor is at temperature $T_S=6\,$K. $-G_{hL}$ is plotted in panel (b) in order for a direct comparison between $-G_{hL}$ and $G_{hR}G_{hR}$. }
\label{fig3}
\end{figure*}

	Since there is no voltage bias applied in each lead, Andreev reflection process is absent, and the electric current in lead $\alpha$ only originates from quasi-particle tunneling and has the expression \cite{NS1, NS2},
\begin{equation}
I_{\alpha} =\frac{2q}{\hbar} \int_{-\infty}^{\infty} \frac{dE}{2\pi} T_{\alpha}^{qp}(E) \left[ f_{\alpha}(E)-f_S(E) \right] .
\end{equation}
where $q$ is the elementary positive charge. 
The electric current in the superconducting lead is thus $I_S=-(I_L+I_R)$. The heat current in three terminals are all carried by electric current in the normal-superconductor junction. This is distinguished from the usual thermal transistors that have been proposed \cite{TT_Li1, TT_Li2, TT_11, TT_NF, TT_FF1, TT_FF2, QTT, TT_Rafa, TT_Coulomb, TT_Ren1, TT_Ren2}. In analogy to the definition of heat amplification factor, we can define a Seebeck amplification factor to characterize the amplification of electric current in a thermal transistor with the form: 
\begin{equation}
\beta_{sb} = \left| \frac{\partial I_L}{\partial I_S} \right| 
=\left| \frac{G_{TL}}{G_{TL}+G_{TR}} \right| .
\end{equation}
$G_{T\alpha}= - \partial I_\alpha /\partial T_S |_{T_\alpha ={\rm const}}$ are the differential thermoelectric conductance at lead $\alpha$ with respect to changing $T_S$.  Similarly to the case of heat amplification, the sufficient and necessary condition for Seebeck amplification factor larger than one is $G_{TL} G_{TR} <0$ and $|G_{TR}| < 2|G_{TL}|$.

\section{Numerical results}  \label{sec:III}
In this section, we provide numerical results regarding tuning amplification factor by changing base temperature and QD levels. We choose the critical temperature of the superconductor to be $T_c = 10\,$K, zero-temperature superconducting gap as$\Delta_0 = 3.0k_B T_c \approx 2.59\,$meV, and Dynes broadening parameter as $\eta_s = 0.02\Delta_0$. The parameter $a$ in the gap formula, equation~\eqref{gap}, is set as $a=0.953$. \cite{gap2} The coupling strengths between the two quantum dots and the three terminals are $\Gamma_L =\Gamma_R = 3.5\,$meV, and $\Gamma_S = 2\,$meV. The left and right normal terminals are maintained at temperatures $T_L=17\,$K and $T_R=3\,$K, respectively. The Coulomb interaction strength is set to be $U=2.0\Delta_0$.

	In Figure~\ref{fig2}, we show the thermal conductance in both left and right normal terminal [panel (a)] and the corresponding heat amplification factor [panel (b)], by varying superconductor's temperature within $T_S \in (T_R, T_c)$. Quantum dot levels are $\epsilon_1 =\epsilon_2 = 3\,$meV. In panel (a), we observe that the NDTC effect for the left terminal happens in the region $T_S \in (3, 6.8)\,$K, and NDTC disappears when the temperature bias between the left and base terminal becomes larger. There is no NDTC for the right terminal as explained in Sec.~\ref{sec:II}. Heat amplification factor $\beta_h^L$ is larger than one only when $T_S \in (3, 6.1)\,$K. Even though there is NDTC effect in the region $T_S \in (6.1, 6.8)\,$K, heat amplification is absent due to that the condition $|G_{hR}|<2|G_{hL}|$ is not satisfied.  Near $T_S=5.7\,$K, heat amplification factor becomes huge and even approaches infinity, this is because $G_{hL}+G_{hR}$ almost vanishes. Hence by tuning the gate temperature, a large amplification factor can be achieved.

	In Figure~\ref{fig3}, heat currents $J_\alpha$ and thermal conductance $G_{h\alpha}$ with respect to the corresponding QD levels are plotted in panel (a) and (b), respectively. Since there is no interaction between the two quantum dots, the left and right heat currents are not influenced by tuning the quantum dot level on the other side, and this also facilitates the tuning of transmission coefficients. One should pay attention that we plot $-G_{hL}$ in panel (b) in order for a direct comparison between $-G_{hL}$ and $G_{hR}$. The magnitudes of heat currents in the normal leads reach their maxima above the zero-temperature gap $\Delta_0$ where the quasi-particle states is at maximum. As indicated by the shadow in panel (b), there are lots of space to tune the QD levels for vanishing $G_{hL}+G_{hR}$, this tells us that a huge heat amplification factor can be easily achieved by changing QD levels. One has much more space to realize huge heat amplification by tuning QD levels compared to tuning the base temperature.

\section{Conclusion}  \label{sec:IV}
We have studied heat amplification in a temperature driven normal metal-superconductor-normal metal three-terminal junction wherein the superconductor terminal acts as the base. Heat amplification requires NDTC to occur only on one side of the junction. Due to the competition between the superconducting energy gap and Fermi-Dirac distribution difference by varying base temperature, NDTC can possibly occur in a normal-superconductor diode when the superconductor side is cooler. If the normal side is warmer, NDTC effect can never happen. The use of quantum dots facilitates the tuning of the heat amplification factor, which can approach infinity.

In this work, we have considered the case that the temperatures of the normal leads are not influenced with changing superconductor's temperature. For possible realistic consideration, we assume that the temperatures of both the left and right lead increase (decrease) slightly with increasing (decreasing) the temperature of superconducting lead $T_S$. With increasing $T_S$, we have increased heat current in the left lead ($J_L$) due to NDTC effect compared to the non-influenced $T_L$. Meanwhile, $-J_R$, which is positive, decreases. From the definition of heat amplification, equation~\eqref{define}, one can see that the heat amplification is decreased slightly compared to the case of non-influenced temperatures in the normal leads.
With decreasing $T_S$, both $J_L$ and $J_R$ decrease, so that the heat amplification is increased slightly.  

One can alternatively use high-temperature superconductor or non-conventional superconductor as the base, then the thermal transistor based on the set-up studied here can work under higher temperatures. The setup presented in this work is reminiscent of a Cooper pair splitter \cite{CPS1, CPS-exp1, CPS-exp2, CPS-exp3, CPS-Yeyati1, CPS-Yeyati2}. There is no interaction (Coulomb interaction or direct coupling) between the two quantum dots in our case. Otherwise, elastic co-tunneling and crossed Andreev reflection processes would exist. There is still thermal transistor effect in presence of interactions between quantum dots, since superconducting energy gap changing due to temperature has minor effect on both the elastic tunneling and crossed Andreev reflection processes. 

\begin{acknowledgements}
G.T. and J.S.W. acknowledge the financial support from RSB funded RF scheme (Grant No. R-144-000-402-114). 
\end{acknowledgements}

\section*{Author contribution statement}
Gaomin Tang conceived the idea and prepared the manuscript. All the authors were involved in the discussion.

\appendix

\section*{APPENDIX: Calculating Green's functions self-consistently}
The retarded and advanced Green's functions ${\bf G}_m^{r/a}(E)$ can be evaluated self-consistently through \cite{Pablo, CPS-Yeyati1}:
\begin{equation}
{\bf G}_m^r (E) = [{\bf G}_m^a (E)]^{\dag} = (g_m^{-1} - \bm{\Sigma}^r_\alpha - \bm{\Sigma}^r_S )^{-1}, 
\end{equation}
with
\begin{equation}
g_m^{-1} = (E - h_m) \left[ 1+\hat{U}(E-h_m-\hat{U})^{-1}\hat{A}_m \right],
\end{equation}
where $h_m = {\rm diag}[\epsilon_m, -\epsilon_m]$, $\hat{U}={\rm diag}[U, -U]$ and $\alpha=L (R)$ for $m=1(2)$. The particle occupation matrix $\hat{A}_m$ is connected with the lesser Green's function via 
\begin{equation}
\hat{A}_m = -i\int \frac{dE}{2\pi} {\bf G}_m^< (E) .
\end{equation}
The lesser Green's function is obtained from Keldysh equation
\begin{equation}
{\bf G}_m^< (E) = {\bf G}_m^r (E) \left[ \bm{\Sigma}^<_\alpha(E)+ \bm{\Sigma}^<_S(E) \right] {\bf G}_m^a (E)
\end{equation}
Since no voltage bias is applied, the lesser self-energies can be obtained from
\begin{equation}
\bm{\Sigma}^<_{\alpha /S}(E) = -2if_{\alpha /S}(E)\ {\rm Im}[\bm{\Sigma}^r_{\alpha /S}(E)] .
\end{equation}

\end{document}